\documentstyle[aps,preprint]{revtex}

\begin{document}
\draft
\title{
 Non-magnetic impurities in two- and three-dimensional Heisenberg antiferromagnets
}
\author{Tai-Kai Ng}
\address{
Department of Physics,
Hong Kong University of Science and Technology,\\
Clear Water Bay Road,
Kowloon, Hong Kong
}
\date{ \today }
\maketitle
\begin{abstract}
  In this paper we study in a large-S expansion effects
of substituting spins by non-magnetic impurities in two- and
three- dimensional Heisenberg antiferromagnets in a weak
magnetic field. In particular, we demonstrate a novel mechanism 
where magnetic moments are induced around non-magnetic impurities 
when magnetic field is present. As a result, Curie-type behaviour
in magnetic susceptibility can be observed well below the
Neel temperature, in agreement with what is being observed in
$La_2Cu_{1-x}Zn_{x}O_4$ and $Sr(Cu_{1-x}Zn_x)_2O_3$ compounds.
\end{abstract}


\narrowtext

 Recently, there has been increasing interests in the effects of 
substituting $Cu$ by non-magnetic ions ($Zn$) in high-$T_c$ cuprates,
where interesting effects were discovered for compounds in the
underdoped regime\cite{uchida}. In particular, magnetic susceptibility 
measurement\cite{ex2} and other experiments\cite{ex3} seem to 
indicate that local magnetic moments are generated as $Cu$ ions 
were replaced by $Zn$, both in the underdoped high-$T_c$ Yttrium
and Bismuth compounds\cite{ex2,ex3}, and for {\em undoped} 
Lanthanum compound \cite{lazn}. Based on RVB theories of 
the $t-J$ model, Nagaosa and Ng\cite{nn} have constructed explanations 
on how local magnetic moments can be generated in these compounds 
in the disordered phase of two-dimensional antiferromagnets. 
Their theory is believed to be applicable to the underdoped regime
of high-$T_c$ compounds. However the theory is not applicable to the 
$La$ compound where Curie-type behaviour is observed in magnetic 
susceptibility at temperatures well below the Neel temperature where
long-range antiferromagnetic order exists. More recently, similar
experiment on the two-leg ladder compound $Sr(Cu_{1-x}Zn_x)_2O_3$
also indicates formation of local magnetic moments in the spin-gap
phase as $Cu$ ions are replaced by $Zn$ ions\cite{az1}. 
The compound was found to
order antiferromagnetically at low temperature at $x\sim0.07$
\cite{az2}. Surprisingly, Curie behaviour in magnetic susceptibility
remains even {\em below} the Neel-ordering temperature\cite{az2},
indicating that formation of magnetic moments around non-magnetic
impurities is a general phenomenon in Neel state of quantum
antiferromagnets.

   In the presence of long-range antiferromagnetic order, the
low temperature properties of clean antiferromagnets can be understood
at least qualitatively using a semi-classical theory ($1/S$ expansion).
In this paper we shall generalize this approach to include the 
effects of non-magnetic impurities in the presence of a uniform magnetic field. 
Notice that in the absence of magnetic field, a similar study has 
been performed by Bulut {\em et. al.}\cite{bulut} where no 'free' 
magnetic moments were found to be induced around 
non-magnetic impurities. We shall show that local magnetic
moments can be induced by non-magnetic impurities once
magnetic field is applied on the system through a novel
mechanism. Notice that strictly speaking, thermal fluctuations
destroy long-range magnetic order at two dimension at any finite
temperature. Thus our analysis at two dimension can only be applied to
layered systems like Lanthanum cuprates where the magnetic
coupling between different layers is much weaker than intra-layer
coupling.

 To begin with, we first consider the zero temperature 
problem of classical spins interacting antiferromagnetically 
under a uniform magnetic field B, with a single non-magnetic impurity 
replacing spin at site $i_0$. We shall consider the Neel state
in the absence of magnetic field 
to be ordered in the $z$-direction, with the uniform magnetic
field $B$ applied in the $+x$-direction. The classical
ground state in the absence of magnetic field has spins all
pointing in $+z$ direction for spins in $A$-sublattice and spins 
all pointing in $-z$ direction for spins in $B$-sublattice. In
the presence of magnetic field, the spins tilt to the
$+x$-direction to minimize the magnetic energy. Let $\theta_i^{A}$
be the angle tilted away from the $z$-axis for spin on site
$i$ on $A$-sublattice and $\theta^{B}_j$ be the corresponding angle
for spin on site $j$ on $B$-sublattice. It is easy to show that
the classical energy $E_{cl}$ is given in the limit when $\theta^{A(B)}$
are small (weak $B$-field limit) by

\begin{equation}
\label{Ecl}
E_{cl}/S^2=-\sum_{<i\neq{i_0},j>}(1-{(\theta^{A}_i+\theta^B_j)^2 \over 2})
-B'\sum_{i\neq{i_0}}\theta^A_i-B'\sum_{j}\theta^B_j,
\end{equation}
where $<i\neq{i_0},j>$ are nearest neighbor sites in the square 
(cubic) lattice excluding contributions from the non-magnetic 
impurity at site $i_0$ and $B'=g\mu_BB/S$. Notice that the only
effect of non-magnetic impurity is to remove the spin at site $i_0$
in our theory. We have also set the spin-coupling $J=1$ in Eq.\ 
(\ref{Ecl}). Notice that the energy expression for $E_{cl}$ is
valid only to order $O(B^2)$. 
  
  Next we consider the continuum limit of the energy expression
\ (\ref{Ecl}). Introducing symmetric and antisymmetric angle
variables 
\begin{eqnarray}
\label{sasa}
\theta_s(\vec{x}) & = & S(\theta^A(\vec{x})+\theta^B(\vec{x})),  \nonumber \\
\theta_a(\vec{x}) & = & {S\over2}(\theta^A(\vec{x})-\theta^B(\vec{x})),
\end{eqnarray}
we obtain after some straightforward algebra
\begin{eqnarray}
\label{Econ}
E_{cl} & = & \int{d^dx\over{a_o^d}}\left[(d)\theta_s(\vec{x})^2-
B'S\theta_s(\vec{x})+(\nabla\theta_a(\vec{x}))^2
+{\theta(x_<)\over{v_a}}\left(\theta_s(\vec{x})
\hat{a_o}.\nabla\theta_a(\vec{x})-\theta_a(\vec{x})\hat{a_o}.\nabla
\theta_s(\vec{x})\right)\right] \nonumber \\
 & & + E_0
\end{eqnarray}
where $E_0=-NS^2d/2$, $N =$ number of sites in the system, 
$a_o$ is lattice spacing and $\hat{a_o}$ is a radial vector pointing
away from $i_0$. $v_a$ is the volume enclosed by the sphere
(circle in 2D) with radius $a_o$. $x_<=a_o-|\vec{x}-\vec{x_0}|$.
For an impurity located at  $B-$ sublattice, 
$\hat{a_o}\rightarrow-\hat{a_o}$.

   The various terms appearing in Eq.\ (\ref{Econ}) can be understood
rather easily. First of all, in the limit $a_o\rightarrow0$, $x_i
\rightarrow{x_j}$ in Eq.\ (\ref{Ecl}) and the only contribution to
$E_{cl}$ would be the first two $\theta_s$ terms in Eq.\ (\ref{Econ}).
It is also clear
that in the limit $a_o\rightarrow0$, terms
proportional to $\theta_a$ or $\theta_a^2$ do not appear in 
$E_{cl}$, and the only contribution from $\theta_a$ can 
appear as $(\nabla\theta_a)^2$ only. Appearance of
non-magnetic impurity at site $i_0$ breaks the symmetry between $A-$
and $B-$ sublattices and introduces local coupling between the 
$\theta_s$ and $\theta_a$ fields. Notice that additional coupling 
between $\theta_a$ and $\theta_s$ fields will appear if we 
take into account higher order terms in $B$ in our energy expression $E_{cl}$.
However, in the absence of impurities, these terms do not break
the symmetry between $A-$ and $B-$ sublattices or the symmetry of
interchanging $\theta^A$ and $\theta^B$ fields. 

  Minimizing $E_{cl}$ with respect to the $\theta_s$ field we obtain
\begin{mathletters}
\label{sol}
\begin{equation}
\theta_s(\vec{x})={1\over2d}\left(B'S-{2\theta(x_<)\over{v}_a}
\hat{a_o}.\nabla\theta_a(\vec{x})\right).
\label{sols}
\end{equation}
Putting $\theta_s$ back into $E_{cl}$ and minimizing with respect to
$\theta_a$ field, we obtain for $|\vec{x}-\vec{x_0}|\geq{a_o}$,
\begin{equation}
\label{sola}
\nabla^2\theta_a(\vec{x})={2B'Sa_o\over{d}v_a}\delta^d(|\vec{x}-\vec{x_0}|-a_o).
\end{equation}
\end{mathletters}

   Notice that in the presence of magnetic field, the non-magnetic
impurity acts as a source term for the $\theta_a$ field of strength
$2B'S$ in the limit $a_o\rightarrow0$. As a result, 
$\theta_a(\vec{x})\sim(2B'S)ln(|\vec{x}-\vec{x_0}|)$ in two dimension,
and the corresponding 'electric field' energy cost
$\sim\int{d}^2x(\nabla\theta_a)^2$ diverges logarithmically as the
size of system goes to infinity.

    The divergence of magnetic energy to order $O(B^2)$ indicates
that the (classical) response of a non-magnetic impurity to external
magnetic field is intrinsically {\em non-linear}\cite{nlc} and 
suggests that quantum effect may play an important role in determining
the correct response of non-magnetic impurities to external
magnetic field. In the following we shall derive in the continuum
limit, the $1/S$ (spinwave) Lagrangian in the presence of background 
$\theta_s$ and $\theta_a$ fields, and shall show that the 
infra-red divergence in classical energy can be cured by
formation of local magnetic moment around non-magnetic
impurity once quantum effects are considered.

   To derive the spinwave Hamiltonian in the presence of magnetic 
field, we rotate our co-ordinate system locally on each site such
that the local $z-$axis is always along the 'classical' spin
direction. In this co-ordinate system, the Hamiltonian becomes

\begin{eqnarray}
H & = & \sum_{<i\neq{i_0},j>}\left[cos(\theta_i+\theta_j)\left(S^{(z)}_i
S^{(z)}_j+S^{(x)}_iS^{(x)}_j\right)+S^{(y)}_iS^{(y)}_j+
sin(\theta_i+\theta_j)\left(S^{(z)}_iS^{(x)}_j-S^{(x)}_iS^{(z)}_j\right)
\right] \nonumber  \\
 & & -g\mu_BB\sum_{i\neq{i_0}}(S^{(z)}_isin\theta_i+S^{(x)}_icos\theta_i)
-g\mu_BB\sum_j(cos\theta_jS^{(x)}_j-sin\theta_jS^{(z)}_j).
\end{eqnarray}
The Hamiltonian can be rewritten in Schwinger-boson representation of
spins in the usual way. In the large-$S$ limit, we may write
\[
Z^A_{i\uparrow}=\bar{Z}^A_{i\uparrow}= 
Z^B_{j\downarrow}=\bar{Z}^B_{j\downarrow}=\sqrt(2S), 
\]
where $\bar{Z}(Z)^{\alpha}_{i\sigma}$'s are spin $\sigma$ 
Schwinger-boson creation (annihilation) operators for site $i$ on
sublattice $\alpha$. Expanding the Hamiltonian to order $O(S)$, we
obtain to order $O(B^2)$, 
\[
H=E_{cl}+H_{1/S},  \]
where
\begin{eqnarray}
\label{hspinwave}
H_{1/S} & = & S\sum_{<i\neq{i_0},j>}\left[(1-{(\theta_i+\theta_j)^2\over2})
\left(\bar{Z}^A_{i\downarrow}Z^A_{i\downarrow}+\bar{Z}^B_{j\uparrow}
Z^B_{j\uparrow}\right)+(1-{(\theta_i+\theta_j)^2\over4})\left(
Z^A_{i\downarrow}Z^B_{j\uparrow}+\bar{Z}^A_{i\downarrow}\bar{Z}^B_
{j\uparrow}\right)\right] \nonumber \\
 & & +g\mu_BB\sum_{i\neq{i_0}}(\theta_i)\bar{Z}_{i\downarrow}^AZ^A_{i\downarrow}
+g\mu_BB\sum_{j}(\theta_j)\bar{Z}^B_{j\uparrow}Z^B_{j\uparrow}.
\end{eqnarray}
 
   To further analyse our system we again go to the continuum limit.
Introducing symmetric and anti-symmetric boson fields
\begin{eqnarray}
\label{bosons}
\phi(\vec{x})={1\over\sqrt2}(Z^A_{\downarrow}(\vec{x})-
\bar{Z}^B_{\uparrow}(\vec{x})), \nonumber  \\
\pi(\vec{x})={1\over\sqrt2}(Z^A_{\downarrow}(\vec{x})+
\bar{Z}^B_{\uparrow}(\vec{x})),
\end{eqnarray}
and integrating out the $\pi(\vec{x})$ field, we obtain in the
continuum limit an effective Lagrangain for $\phi(\vec{x})$
field\cite{rs,ng95}. Details of the technique can be found in
ref.\cite{rs} and we shall not repeat them here. 
For $\vec{x}$ away from the impurity site
($a_o<|\vec{x}-\vec{x_0}|$), we obtain in imaginary time

\begin{equation}
\label{leff}
{\em L}_{1/S}\sim\int{d}\tau\int{d}^dx{1\over4dS}\left[({\partial\over
\partial\tau}-e\theta_a)\phi^+({\partial\over\partial\tau}+e\theta_a)\phi
+m^2\phi^+\phi+c^2|\nabla\phi|^2\right],
\end{equation}
where $e=B'$ and $m^2=S^2B'^2/2$. $c^2\sim4d(JS)^2$ is the spinwave 
velocity. We have set $\theta_s(\vec{x})=B'S/2d$
and have neglected $(\nabla\theta_a)^2$ terms in deriving ${\em L}_{1/S}$.
The later is of higher order $(O(1/S))$ compared with the corresponding 
term in $E_{cl}$. The most striking feature of the effective Lagrangian is
that the $\theta_a$ field now appears as the $\tau$-component of a U(1)
gauge field coupling to a charge boson field $\phi$, 
with the dynamics of the $\theta_a$ field governed by $E_{cl}$. 
In particular, in the presence of
$B$ field, the non-magnetic impurity appears as electric charge 
generating a static electric field coupling to the charge bosons.
Notice that the effective charge of the non-magnetic impurity changes
sign when it is moved from one sublattice to another, indicating that
the 'sign' of the charge is in fact a sublattice index\cite{rs,ng95}.

    The logarithmic divergence in 'electric field' energy in presence
of non-magnetic impurity in $E_{cl}$ in two dimension will be removed
if bosons of opposite 'electric' charge are nucleated from vacuum to 
screen the effective electric field, forming effective
local magnetic moments around the impurity. The number of
bosons nucleated from vacuum is $\sim2S$, as can be seen easily
by counting the number of 'charges' carried by the non-magnetic
impurity. The magnitude of magnetic moment formed around the
impurity is thus $\sim{S}$. The energy 'cost' for nucleating
the bosons can be computed by solving the corresponding Schr$\ddot{o}$dinger
equation for charge bosons moving in scalar potential of external charge
of magnitude $2Se$\cite{se}. To capture qualitatively the
physics, we estimate the energy 'cost' by using a variational
wavefunction $\psi(\vec{r})\sim{A}e^{-|\vec{r}|/\xi_1}$, where
$\xi_1$ is determined variationally. Minimizing the energy, we
find that $\xi_1$ is of the order $\xi_1\sim
(JSc^2/(g\mu_BB)^3)^{1/2})$. Notice that $\xi_1\rightarrow\infty$
as $B\rightarrow0$ or $S\rightarrow\infty$, indicating that
the formation of magnetic moment around non-magnetic impurity
is a non-perturbative quantum effect which cannot be captured
by usual spinwave theory. The energy 'cost' for nucleating the
bosons is of order $E_1\sim(2S)(m+2SB'^2ln(\xi_1/a_o))$.

   Next we discuss the situation when there is finite concentration
of impurities $n$ randomly distributed in the system, which is the 
case of experimental interest. In our continuum theory, the presence of
finite concentration of random impurities is equivalent to putting finite
concentration of 'charges' with magnitude $2SB'$ and random sign
in our effective charge boson system. The behaviour of the system
is very different from the one impurity case, since now 
electric fields originating from opposite charges will cancel, 
with remaining 'electric field' energy of order $\sim{n}E_e$, 
where $E_e\sim(2SB')^2ln(l/a_o)$ on average, $l\sim{n^{-1/d}}$ 
is the average distance between non-magnetic impurities. For
$E_e>E_1\sim2Sm$ or equivalently
$l>>\xi_o\sim{a_o}e^{JS/(2g\mu_BB)}$, it is energetically
favourable to nucleate charge bosons from vacuum to screen the 
electric field, and magnetic moments will be formed around
non-magnetic impurities. The magnetic moments couple to each other
weakly with $J_{eff}\sim{J}e^{-l/\xi_1}$ implying that 
Curie behaviour will be present in magnetic susceptibility down
to very low temperature $\sim{J_{eff}}$. As concentration of
non-magnetic impurities increases, $l$ decreases until $l\sim\xi_o$.
At this point a transition occurs where it becomes energetically 
unfavourable to nucleate bosons from vacuum to screen the electric
field, i.e. there will be no magnetic moments forming around
non-magnetic impurities at zero temperature when $l\leq\xi_o$.
The boson-non-magnetic impurity bound states become excited states
of the spin system!

  The boson-non-magnetic impurity bound states can still be
observed as effective local magnetic moments at finite temperature $T\geq
{m}\sim{g\mu_BB}$ if $E_e+2Sm>E_1$ (corresponding roughly to
$\xi_1<l$). At this energy range, the boson-impurity bound state
appears as stable excited state of the spin system with excitation energy
$\sim(E_1-E_e)/2S<m$. At finite temperature $T\geq{m}$, these states 
will be occupied by thermally excited bosons, forming effective 
free magnetic moments around non-magnetic impurities.
It is important to emphasize that the formation of excited boson
bound state around non-magnetic impurity is only possible 
when there is a gap  $m$ in the boson excitation spectrum. In the absence 
of non-magnetic impurity, a direct computation of the spinwave 
spectrum indicates that the spinwave spectrum splits into 
two branches at small momentum $q$, with one branch carrying a gap larger 
than $m$ in our effective Lagrangian, and the other branch gapless as $q
\rightarrow0$. The contribution from the splitting of the two branches
to $H_{1/S}$ cancels to order $O(B^2)$, leading to our effective
Lagrangian with only massive bosons. The splitting of the boson
spectrum reappears if we include terms to order $O(B^4)$ in $H_{1/S}$,
and the boson bound state around non-magnetic impurity will cease
to be good eigenstate of the effective Lagrangian (the bound state
turn into resonant state). However, the decay rate of the
resonant state $\sim{B}^4$ is small compared with the energy of the 
resonant state $\sim{B}^2$ in the limit of weak magnetic field, 
indicating that the physical picture of boson bound state around 
non-magnetic impurity is still a good approximation to
the system in weak field limit.

  As concentration of impurities increases and distances between
impurities decreases further, the overlap between boson bound state
wavefunctions at different impurity sites increases and a boson impurity band 
is formed (recall that the size of the boson bound state
wavefunction is of order $\xi_1$). For $l\leq\xi_1$, the bandwidth
$W$ of this impurity band is of order $\sim(\sqrt{m^2+(c/l)^2}-m)$ and the 
effective impurity potential strength is of order $v\sim2SB'^2/(1/l)^2$.
We find that the screening of effective impurity potential by 
charged bosons is weak and can be neglected. Estimating the scattering
life time $\tau$ we find that bosons with energy
$E\leq{E}_l\sim{c}l/\xi_1^2$ are strongly localized in this impurity
band using the criteria $E\tau\leq1$. Notice that the number of localized 
boson states $\sim{n}^{-1} $ and decreases as $n$ increases in this
regime. The magnetic susceptibility will show Curie behaviour at
temperature above $T_l\sim{E}_l/k_B$. For $T\leq{T}_l$, the number of 
effective magnetic moments contributing to magnetic susceptibility 
decreases as temperature lowers. 
Roughly speaking, the contribution from non-magnetic impurities 
to the magnetic susceptibility can be described by a magnetic field 
and temperature dependent density of local moments $n(B,T)$ which 
has the following properties: (i) $n(B,T)\sim0$ at the region 
$g\mu_BB>>k_BT$, (ii) $n(B,T)\sim{n}$ for $l\geq\xi_1$, and
$\sim(n\xi_1^4)^{-1}$ for $l\leq\xi_1$ at the region $k_BT>>g\mu_BB+E_l$, 
and is smoothly interpolating between the two regions. As
temperature $T\rightarrow0, n(B,T)\rightarrow0$ and the contribution
to magnetic susceptibility from impurities is of order
$\delta\chi\sim4S^2nln(l/a_o)$. 

   Our theory can be extended to three dimension in a
straightforward way. We find that in the weak-field limit, no
magnetic moment is formed even in the limit of one single impurity
at zero temperature because of absence of infra-red divergence associated with
Coulomb potential in $3D$. However, bound states of bosons can be formed 
at $T\geq{g}\mu_BB$ as in two dimensional case, and can
still result in Curie-type behaviour in magnetic susceptibility for
$l>>\xi_1\sim{(c/g\mu_BB)^3}$. As concentration of impurities
increases further, the boson bound states turn into impurity band
and the effective number of 'free' magnetic moments $n(B,T)$ decreases 
when temperature is lower than the boson impurity band bandwidth 
as in two dimension.

  Summarizing, in this paper we have carried out an analysis of the
effects of replacing spins by non-magnetic impurities in 
Heisenberg antiferromagnets in magnetic field in two- and three- 
dimensions in a semi-classical $1/S$ expansion. 
We find within a continuum approximation that
magnetic moments will form around non-magnetic impurities
once magnetic field is put on the system, resulting in
Curie-type behaviour in magnetic susceptibility when
concentration of impurities is not too high, and temperature
is not too low. Notice that our theory predicts that formation
of local magnetic moments in presence of external magnetic field
and non-magnetic impurities is a general behaviour of ordered
quantum antiferromagnets and is not restricted to particular materials.
The theory explains the observation of
Curie-type behaviour in magnetic susceptibility in 
$La_2Cu_{1-x}Zn_xO_4$ and $Sr(Cu_{1-x}Zn_x)_2O_3$ compounds at temperature
well below Neel temperature. In particular,
the vanishing of local moments at $T\leq{g}\mu_BB$ is a
theoretical prediction which can be tested experimentally.

 I thank G. Aeppli, M. Ma and M. Azuma for very useful discussions and the hospitality 
of Lucent Technologies, Bell Labs. for where part of this work is 
done. The work is partially supported by Hong Kong UGC Grant. UST636/94P.


\begin{references}
\bibitem{nlc} It can be shown that the logarithmic divergence in classical
 linear response is removed once higher order terms in $\theta_s$ and
 $\theta_a$ are included in the energy expression. The resulting classical 
 energy is of order $B^2ln(SJ/\mu_BB)$.
\bibitem{se} The Schr$\ddot{o}$dinger equation for charged boson in the
 effective scalar potential of a external point charge is $
(E-V(r))^2\psi(\vec{r})=(c^2\nabla^2+m^2)\psi(\vec{r})$, where $V(r)\sim2SB'^2ln(r/a_o)$.
\end{references}
\end{document}